\documentclass[twocolumn,tightenlines,showpacs,aps]{revtex4}
\usepackage{graphicx}

\begin{document}

\title{On the Experimental Estimation of Surface Enhanced Raman Scattering (SERS) Cross Sections by Vibrational Pumping}
\author{R. C. Maher} \email{Robert.Maher@imperial.ac.uk}
\author{L. F. Cohen}
\affiliation{The Blackett Laboratory, Imperial College London\\
Prince Consort Road, London SW7 2BW, United Kingdom}
\author{E. C. Le Ru}
\author{P. G. Etchegoin} \email{Pablo.Etchegoin@vuw.ac.nz}
\affiliation{The MacDiarmid Institute for Advanced Materials and Nanotechnology\\
School of Chemical and Physical Sciences, Victoria University of
Wellington,  PO Box 600 Wellington, New Zealand}

\date{\today}

\begin{abstract}
We present an in-depth analysis of the experimental estimation of
cross sections in Surface Enhanced Raman Scattering (SERS) by
vibrational pumping. The paper highlights the advantages and
disadvantages of the technique, pinpoints the main aspects and
limitations, and provides the underlying physical concepts to
interpret the experimental results. Examples for several commonly
used SERS probes are given, and a discussion on future possible
developments is also presented.
\end{abstract}


\maketitle

\section{Introduction}

In the last decade Surface Enhanced Raman Scattering
(SERS)\cite{Moskovits} has made rapid progress towards
applications.  With a sensitivity rivaling fluorescence in some
cases, and a much higher structural specificity, SERS is a highly
attractive technique, being developed simultaneously with the
field of plasmonics. Progress towards different uses of SERS in
practical applications has been steady. New substrates including:
arrays of inverted pyramids\cite{tuning}, silver pillar and
torroid arrays\cite{cheerios}, adaptive silver
films\cite{adaptive}, and metallic nano-shells\cite{nanoshells},
have been demonstrated. In addition, many molecules relevant to a
myriad of applications such as glucose\cite{glucose},
proteins\cite{proteins}, DNA\cite{DNA}, a wide range of medicinal
drugs\cite{antimalarial-drug, cancer-drug,
leukemia-drug-characterisation}, and substances for forensic
science\cite{explosives}, have been characterized. On the other
hand, the understanding of some fundamental aspects of the
phenomenon is still incomplete, and in some cases controversial.
Although a full understanding is sometimes not essential for the
development of applications, there can be no denying the fact
that better comprehension is desirable.

Optical pumping of vibrational modes was first suggested in 1996
based on the observed dependence of the anti-Stokes/Stokes (aS/S)
ratio with incident laser power\cite{Kneipp1}, and it was
suggested that this could be a possible tool to estimate SERS
cross sections. However, the interpretation of the experimental
results have been the subject of considerable debate in the
literature, with many authors simply denying its existence and
attributing the experimental observations to either laser-heating,
resonance effects, or combinations
thereof\cite{Brolo,Haslett,HPDef,Teredesai}. Over a series of
previous papers we have suggested the investigation of the
aS/S-ratio as a function of temperature $(T)$ as an alternative
method for the observation of
pumping\cite{RobFaraday,RobJPCB,RobJPCB2}. In a recent paper, we
provided definitive evidence for the existence of vibrational
pumping under SERS conditions thereby bringing to conclusion one
aspect of this complex problem\cite{RobJPCB2}. Here we address
another aspect, namely the problems associated with estimating the
SERS cross section using this method, and  we extend our analysis
to include different analytes and laser excitations. We also
highlight several outstanding issues which were not considered
previously\cite{RobJPCB2}. Obtaining good estimates of SERS cross
sections is in general an extremely hard problem and has been a
longstanding ambition of the SERS community for reasons that go
from the purely applied (quantification of signals) to the more
fundamental (comparisons of theoretical electromagnetic
enhancement factors with experiment). Any method that can produce
a standard protocol for the estimation of SERS cross sections is,
accordingly, of great interest and an effort to understand its
principles and limitations is required.

Before we go into the details, we briefly review the necessary
concepts for understanding aS/S-ratios and vibrational pumping in
SERS. The method we are going to present has not been widely used
and therefore we review  the historical background and also its
peculiarities and limitations. The discussion has some natural
overlap with our previous papers\cite{RobFaraday,RobJPCB,RobJPCB2}
but it is presented here for the sake of completeness, the
convenience of the reader, and future reference for forthcoming
work in progress. This will inevitably result in a somewhat
lengthy introductory section where the principles of the method
are laid down. The next sections are fully devoted to this and are
followed by a section at the end with a practical experimental
demonstration of SERS cross section determination for several
standard probes.

\section{A brief description of SERS pumping}

\subsection{The Anti-Stokes/Stokes ratio}

Let us consider the different contributions to the population of a
single vibrational level at temperature $T$ during a SERS
measurement; we can identify two main contributions: $(i)$ the
laser itself which pumps vibrations through Raman processes with a
rate proportional to its intensity $(I_L)$ and to the Raman-Stokes
cross section $(\sigma_{S})$, and $(ii)$ thermal excitation and
relaxation. Vibrations remain in the level with a finite lifetime
$\tau$, which encompasses all possible relaxation mechanisms
--such as intramolecular vibrational relaxation IVR (anharmonic
processes) or external relaxation mechanisms. There are other
possible secondary mechanisms, such as relaxation through an
anti-Stokes Raman processes, or excitation to higher vibrational
levels. It is possible to write rate equations for the detailed
dynamics of vibration populations in such a system \cite{HPDef}.
In the regime of weak pumping (which is the only case considered
here), where the vibrational population remains small, $n \ll 1$,
it is sufficient to consider only the two main mechanisms, and the
rate equation for $n$ can then be written as:
% Gaining this
%relaxation time from the SERS measurement itself is, in fact, one
%of the main limitations of this technique, as we shall show later.
%We mention in passing at this stage that $\tau$ is in general {\it
%not directly obtainable from the width of the peaks which has a
%contribution from population dephasing and inhomogeneous
%broadening too}. We shall come back to this problem later. In the
%interim, let us assume that we know $\tau$. The rate equation for
%the population $n$ of the level can be written as:
\begin{equation}
\frac{dn}{dt} = \frac{\sigma _S I_L}{\hbar \omega_L} +
\frac{\exp(-\hbar \omega_{\nu}/k_BT)}{\tau} - \frac{n}{\tau},
\label{rateEqn}
\end{equation}
where $\sigma_S$ is the Raman-Stokes cross section, $I_L$ the
intensity (power per surface area) of the laser, $\hbar\omega_L$
the energy of an exciting photon ($n_L=I_L/\hbar\omega_L$ is the
number of incident photon per unit time and surface area), and
$\hbar\omega_{\nu}$ the energy of the vibration. The first term on
the right is the number of vibrations per unit time being pumped
into the level by the action of the laser, while the second and
third terms are the contributions of thermal excitation and
population relaxation, respectively. In the steady state $dn/dt =
0$. Generally, $\sigma_S$ is very small, so when $I_L$ is small
the pumping contribution is negligible and the vibrational
population is dominated by thermal effects; i.e. $n =
\exp(-\hbar\omega_v/k_B T)$ is given by a Boltzmann factor. When
$I_L$ is increased (so that there is a significant pumping
contribution to the population of the level) $n$ becomes:
\begin{equation}
n = \frac{\tau\sigma _S I_L}{\hbar \omega_L} + e^{-\hbar
\omega_{\nu}/k_BT}, \label{nSteady}
\end{equation}
where the pumping term is clearly distinguished from the thermal
contribution.

In addition, for an ensemble of $N$ molecules, the Stokes-Raman
signal is given by $I_S=N\sigma_S I_L$ while the anti-Stokes
signal is given by $I_{aS}=n N \sigma_{aS} I_L$, leading to:
\begin{equation}
I_{aS}=\left(\frac{\tau\sigma _S I_L}{\hbar \omega_L} + e^{-\hbar
\omega_{\nu}/k_BT}\right)N\sigma_{aS}I_L. \label{quadASint}
\end{equation}
Moreover, taking the ratio of the anti-Stokes to the Stokes
intensities $(\rho)$, we have:
\begin{equation}
\rho = \frac{I_{aS}}{I_S} = \frac{\sigma_{aS}}{\sigma_S}n = An,
\end{equation}
where $A$ is the {\it asymmetry factor} (extensively discussed in
Refs. \cite{RobFaraday,RobJPCB,RobJPCB2}). $A$ includes not only
any possible difference between anti-Stokes and Stokes cross
sections arising from resonance effects (due either to plasmon
resonances or to resonant Raman scattering for a resonant
analyte), but also the standard wavelength dependence of Raman
processes $A_{\omega_{\nu}} =
(\omega_L+\omega_{\nu})^4/(\omega_L-\omega_{\nu})^4$. $\rho$ can
be expressed as:
\begin{equation}
\rho = A\left[\frac{\tau\sigma _S I_L}{\hbar \omega_L} + e^{-\hbar
\omega_{\nu}/k_BT}\right]. \label{pumpRatio}
\end{equation}

This simple model shows that the Stokes-Raman signal always
remains linearly dependent on $I_L$. The anti-Stokes signal also
shows a linear dependence with power when pumping is negligible.
But when pumping dominates over thermal effects, then $I_{aS}$
varies quadratically with power while the aS/S-ratio is linearly
dependent on $I_L$.

The original work by Kneipp et. al.\cite{Kneipp1} used rhodamine
6G (RH6G) and crystal violet (CV) (two commonly used SERS active
dyes) under $830\, {\rm nm}$ excitation at room temperature (RT).
The argument for pumping was primarily based on two observations:
(i) the aS/S-ratios were shown to be larger than expected for a
Boltzmann factor, and (ii) the {\it power dependence} of the
signal at RT (where the thermal contribution is likely to
dominate, or be important) was quadratic. This observation of a
non-linear dependence of the anti-Stokes intensities with power
(resulting in a linear dependence in the aS/S-ratio) was thought
to provide strong evidence for vibrational pumping in SERS.
Further measurements on carbon nanotubes\cite{Teredesai} and a DNA-base\cite{KneippAdenine} supported these initial results. The
arguments against this original interpretation are summarized in
the following subsection.

\subsection{Resonance and heating effects}

Haslett et. al.\cite{Haslett} were the first to seriously question
the existence of vibrational pumping as revealed in the original
studies. Extensive measurements were made under similar conditions
with a number of both resonant and non-resonant molecules.  They
observed an anomalous ratio for all the resonant molecules tested
which was independent of power until photo-bleaching occurred. No
anomaly was observed in the case of the non-resonant molecules. It
was concluded that the anomalous ratios observed in the original
papers were the result of ``hidden'' resonances rather than
pumping\cite{MaherResonance}. Brolo et. al.\cite{Brolo} also
concluded that the observed anomalous ratio could be explained by
resonances. These resonances are accounted for by the asymmetry
factor $A$ in the model of the previous section. In the absence of
pumping, the aS/S-ratio is predicted to be:
\begin{equation}
\rho=A\exp \left( -\frac{\hbar\omega_{\nu}}{k_B T} \right).
\label{ratio-corrected}
\end{equation}
It is clear that the $A$ factor can indeed result in anomalous
ratios (different from the Boltzmann factor), even in the absence
of pumping.

Furthermore, it was suggested that the observed power dependence
was not the effect of pumping but rather laser heating. The effect
of heating on $\rho$ in the absence of pumping can be simply
understood by including it in Eq. (\ref{ratio-corrected}). The
real temperature of the probed molecule, $T_h$, may be different
to the nominal temperature $T$ because of heating effects, either
in the SERS substrate or in the molecule itself \cite{HPDef}. By
writing $T_h=T+\Delta T$ and assuming that $\Delta T \ll T$ we can
obtain the corresponding ratio, $\rho_h$, by expanding Eq.
\ref{ratio-corrected} as:
\begin{equation}
\rho_h = A\exp\left(-\frac{\hbar\omega_{\nu}}{k_B
T}\right)\exp\left(\frac{\hbar\omega_{\nu}}{k_B T}\frac{\Delta
T}{T}\right). \label{ratio-X}
\end{equation}
This expression first shows that heating can also result in an
anomalous aS/S ratio, even when $A\approx 1$. Moreover, it can
have an important impact on the power dependence of $\rho$,
because $\Delta T$ should increase with $I_L$. We can assume in a
first approximation that $\Delta T \approx \alpha I_L$, since heat
diffusion and transfer are linear problems. The above expression
shows that laser heating should then result in an exponential
increase of $\rho_h$ with $I_L$. However, in many cases of
interest, the argument in the second exponential is small compared
to 1, and Eq. (\ref{ratio-X}) can then be further expanded to
give:
\begin{equation}
\rho_h = A\exp\left(-\frac{\hbar\omega_{\nu}}{k_B T}\right)\left(1
+ \frac{\hbar\omega_{\nu}}{k_B T}\frac{a}{T}I_L\right).
\label{ratio-XX}
\end{equation}
A linear dependence of $\rho$ with $I_L$ (or equivalently a
quadratic dependence of $I_{aS}$ with $I_L$) {\it can then equally
be the result of conventional heating effects or pumping}. Such an
observation is therefore insufficient to demonstrate the presence
or not of vibrational pumping. We also note that heating effects
can also explain the mode-dependent behavior observed in the
original work \cite{Kneipp1}, as the argument of the second exponential in Eq.
(\ref{ratio-X}) depends on the mode energy $\hbar\omega_{\nu}$.

\subsection{Temperature dependence of the aS/S ratios}

Most studies have concentrated on finding evidence for SERS
pumping at room temperature (RT), in particular by studying the
power dependence of the aS/S ratio. It is true that the huge
enhancements in SERS conditions greatly increases the contribution
from pumping, but may also contribute to an increased heating (in
particular directly in the probe molecules). At room temperature
(RT) and above, the thermal contribution to $\rho$ is relatively
large and in many cases dominates, such that $\rho \approx A
\exp(-\hbar\omega_{\nu}/k_BT)$. The study of vibrational pumping
at RT therefore involves measuring small departures from an
already existing (large) thermal population. Moreover, as the
previous discussion has shown it is also extremely difficult to
distinguish the relative contributions of pumping and heating at
RT.

\begin{figure}
 \centering{
  \includegraphics[width=10cm, height = 8cm]{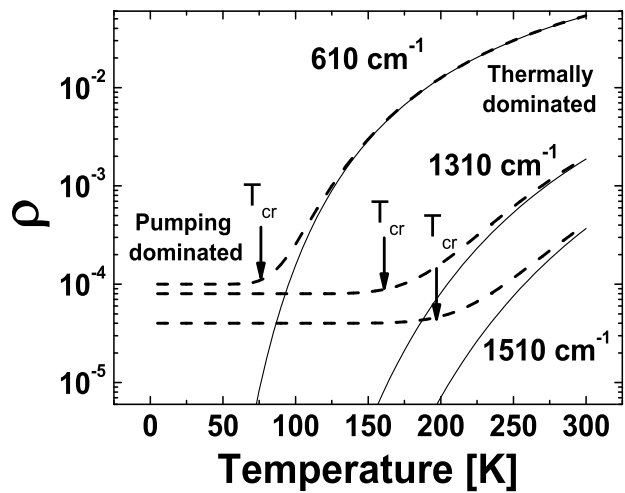}
 }
\caption{Schematic diagram showing the temperature dependence of
the anti-Stokes/Stokes ratio ($\rho$) for several different Raman
modes. We have assumed here $A=1$ for simplicity and have used
three characteristic Raman modes of RH6G for the example. The
solid lines represent the ratio when there is no pumping
contribution (Boltzmann factor). The dashed lines show the result
of including pumping as given by Eq. \ref{pumpRatio}. There are
two regimes, one at high temperature where the thermal
contributions dominate and the ratio is similar to what is
expected if no pumping were present. The second occurs at low
$T$'s where the contribution of pumping becomes significant
relative to the thermal contribution. We refer to the cross-over
point between these two regimes as the $T_{cr}$ (indicated by the
arrows in this figure). $T_{cr}$ generally occurs at higher $T$'s
for higher energy modes due to the different influence of the
exponential factor for different mode energies in Eq.
\ref{pumpRatio}. See text for further details.}
 \label{SchmRatios}
\end{figure}

A more practical approach is to study the temperature dependence
of the aS/S ratios, and in particular the low-temperature regime
where thermal effects are expected to be completely negligible
compared to any other mechanisms, and in particular pumping. The
temperature dependence can be directly studied using the model
presented so far. Figure \ref{SchmRatios} summarizes the different
scenarios for aS/S-ratios as a function of temperature. The solid
lines show the variation of the ratio with $T$ when pumping is
absent; i.e. an exponential decrease as $T$ decreases, following
the Boltzmann factor $\exp(-\hbar\omega_v/k_B T)$. The dashed
lines show the case when pumping is present. At high $T$'s the
ratio is approximately the same as if pumping were absent. This is
the {\it thermally-dominated regime}. As $T$ is decreased, there
is a cross-over to a {\it pumping-dominated regime}, where the
ratio reaches a plateau. We refer to the cross-over temperature
between these two regimes as T$_{cr}$; indicated by arrows in Fig.
\ref{SchmRatios}. The cross-over occurs when pumping and thermal
terms in Eq. (\ref{pumpRatio}) are comparable, which occurs for:
\begin{equation}
k_B T_{cr} \approx \hbar\omega_{\nu}\left[ \ln\left(
\frac{\hbar\omega_L}{\tau \sigma_S I_L} \right) \right]^{-1}
\end{equation}
The cross-over occurs at a larger temperature for higher powers
$I_L$, for which pumping is stronger, but also for higher energy
peaks (with a larger $\hbar\omega_{\nu}$), for which thermal
excitation is weaker.

In the pumping-dominated regime, the aS/S-ratio is constant and
equal to $\rho=A \tau \sigma_S I_L/(\hbar\omega_L)$.
$I_L/(\hbar\omega_L)$ can be estimated to a high degree of accuracy for a
given experimental setup. If $A=1$ (no asymmetry between
$\sigma_{S}$ and $\sigma_{aS}$) we can therefore deduce the
product $\tau \sigma_S$ from the plateau in the aS/S-ratio below
$T_{cr}$. In general, and in particular under SERS conditions,
$A\neq 1$, but its value can be determined from the exponential
dependence in the thermally-dominated regime (above $T_{cr}$). In
practice, a fit of the experimental data over the whole
temperature range with two parameters enables us to determine both
$A$ and $\tau\sigma_S$ \cite{RobJPCB2}. Nevertheless, the
problem of estimating $\tau$ still needs to be circumvented to
determine the SERS cross-section $\sigma_S$ itself.

 By reducing
$T$, the contribution from pumping becomes dominant, making
measurements much easier, and more reliable. It also simplifies
greatly their interpretation. But the main advantage is that it
enables one to unambiguously rule out heating effects as an
alternative explanation. As discussed, the appearance of plateaus
for $T<T_{cr}$ in Fig. \ref{SchmRatios} is a clear
characterization of the pumping-dominated regime. Note that it
does not mean that heating is absent, but simply that its effect
on aS/S-ratios is negligible (relative to the
pumping contribution). Any heating effects may possibly affect the
value of $T_{cr}$, but as long as a plateau is observed,
vibrational pumping must occur. Moreover, the value of $\rho$ in
the plateau region (from which we will infer SERS cross-sections)
is independent of heating.

Studies of the temperature dependence of $\rho$ therefore provide
a more conclusive demonstration of SERS pumping. However, the
technique is not without its drawbacks. In order to extract SERS
cross-sections for the method discussed here there are several
issues that need special attention as described below.

\section{Practical estimation of cross-sections from vibrational pumping}

Several issues need to be considered in the practical
determination of SERS cross sections via vibrational pumping. Not
all of them can be resolved satisfactorily and we have to include some {\it
ad-hoc} approximations in order to obtain estimates of the cross
sections. While there is nothing intrinsically wrong with this, it
is necessary to be aware of the range of validity of the estimates
in order to be able to comprehend the current limitations of the method as proposed.

\subsection{SERS cross-sections}

It is worth mentioning here that in most situations in Raman (and
in SERS) we talk plainly about the cross section $\sigma$ while in
reality there are at least three possible cross sections we can
refer to: $(i)$ radiative, $(ii)$ non-radiative, and $(iii)$
total. In the case of vibrational pumping, for example, the rate
at which vibrations are pumped into the molecule is proportional
to the {\it total} cross section. But some of the re-emitted
photons might not be observable in the far-field because they are
absorbed by plasmon resonances in the vicinity of the probe. This
is what makes the distinction between {\it radiative} and {\it
non-radiative} cross sections. This could lead potentially to
inconsistencies among cross sections for different modes if they
happen at frequencies where the non-radiative contributions are
different. We will not insist on this distinction in the following,
but it is worth keeping in mind that as far as the cross sections
is concerned what we measure is the ``radiative'' part of an
effect (pumping) which depends on the ``total'' (radiative plus
non-radiative) cross section. This distinction could under certain circumstances be important
.

\subsection{Vibrational lifetimes}

We summarize in this subsection the different aspects of the
problem of lifetime estimation for the purpose of obtaining SERS
cross sections.

\begin{itemize}
\item{Ultimately, the observation of a cross-over from a
thermally-dominated to a pumping-dominated aS/S-ratio leads to an
estimation of both the asymmetry factor $A$ and the product $\tau
\sigma_{S}$. An estimation of $\sigma_{S}$ itself requires the
knowledge of $\tau$, as pointed out before. The latter cannot be
directly obtained, in general, from the SERS spectrum potentially leading
to a serious limitation of this technique. In the
first report of cross section estimation via vibrational
pumping\cite{Kneipp1} a lifetime of $\tau\sim 10\, {\rm ps}$ was
simply estimated without any reference to the experimental data.}
\end{itemize}

In order to improve upon this we need to make a series of
assumptions. The validity of these assumptions needs to be
evaluated on a case-by-case basis and the results obtained for the
cross section have to be interpreted within these
approximations/assumptions. Let us first review the problems
associated with trying to extract lifetimes from the SERS spectra.

\begin{itemize}

\item{Raman peaks in molecules have lineshapes which are seldom
pure Lorentzians and have several contributions to their natural
widths; a subject of a longstanding history in
spectroscopy\cite{KaiserRMP}. The observed linewidth has typically
contributions from: $(i)$ groups of Raman modes piled up together
in narrow energy regions; $(ii)$ inhomogeneous broadening, $(iii)$
phase coherence relaxation, and $(iv)$ population relaxation. The
last two are also known as the {\it off-diagonal} (dephasing) and
{\it diagonal} (population) relaxation times in density matrix
formalism\cite{KaiserRMP}, respectively, and are part of the {\it
homogeneous broadening} of the peak. These relaxation times are
different from mode to mode and cannot be easily separated from
the observed linewidth in a plain SERS spectrum. The {\it
population relaxation} is the one we need ($\tau$) for the
estimation of $\sigma_S$.}

\item{The presence of multiple modes contributing to a peak is a
drawback that can be overcome in many situations.  In the case of dyes it is not unusual for several Raman active modes, which are closely spaced in energy and have different cross-sections, to contribute to the observed peaks.  This will naturally result in distortions of the
lineshape and problems with any estimate of the lifetime based on
the width of the peak. This can be overcome by the use of well
characterized vibrations which are relatively isolated from the
others and even (if possible) complementing the information by
density functional theory (DFT) calculations of the Raman
spectra\cite{DFT1,DFT2} to support the selection of those modes
that are the best candidates for a reliable estimation of $\tau$.}

\item{If the broadening is homogeneous\cite{book}, the only
remaining contributions to the linewidth are the diagonal and
off-diagonal relaxation times. Molecular vibrations have typically
very strong anharmonic couplings with other vibrations in the
molecule and contributions to inhomogeneous broadenings are mostly
negligible. Even in SERS experiments where the single molecule
limit is approached\cite{UsJPC} (which would be more sensitive to
inhomogeneous broadening) the full width at half maximum (FWHM),
$\Gamma$, of the peaks does not change by more than $1-1.5$
cm$^{-1}$ in peaks with a typical $\Gamma$ of $\sim 15-20$
cm$^{-1}$. The validity of this assumption (whether inhomogeneous
broadening is important or not) has to be assessed obviously in
each specific situation.}

\item{The separation between diagonal and off-diagonal relaxation
times cannot be achieved by a simple SERS spectrum. It is always
possible to invoke a different type of spectroscopy (time
resolved)\cite{KaiserRMP}, but this may not be feasible in most
situations. One simple possibility (as done in Ref.
\cite{RobJPCB2}) is to estimate the population relaxation by using
the FWHM $(\Gamma)$ of the peaks via $\tau\sim \hbar/\Gamma$; i.e.
essentially ignoring the contribution of the dephasing
(off-diagonal) relaxation to the width. This will produce an
underestimation of $\tau$ and an overestimation of $\sigma_{S}$.
In addition, this typically produces cross-sections for different
peaks which are not entirely consistent with the relative
(integrated) intensities among peaks; thus showing an intrinsic
inconsistency in the estimation of the $\tau$'s (which can be of
the order of a factor of 2 to 10 for the examples we examined). It
is generally difficult to make a simple estimation that will be
valid for all modes. It is quite clear that for the cross sections
to be meaningful (and if they are not affected by different
amounts of non-radiative processes) the relative cross sections
among peaks must be in accordance with the relative integrated
intensities observed in a normal SERS spectrum.}

\item{Population relaxation is caused mainly by the anharmonic
coupling to the ``thermal bath''. This bath includes all other
modes in the molecule as well as those in the solvent or
substrate. Generally speaking, higher energy modes have shorter
lifetimes due to a greater number of possible decay pathways.
Intramolecular anharmonic decays are energy conserving, meaning
that high energy modes have more possibilities for a decay than
low energy ones. Population relaxation is also the only dynamical
process that contributes to the vibrational linewidth in the limit
$T\rightarrow0$.}

\item{It is important to realize that, in principle, {\it if we
find a single peak in the spectrum where the population relaxation
can be gained directly from the FWHM via} $\tau\sim \hbar/\Gamma$,
{\it then all of the cross sections of the other modes follow
immediately through the relative integrated intensities of the
peaks}. When lacking a time resolved experiment to isolate the
contributions properly, {\it the problem comes down to a judicious
choice of a mode where the width is dominated by relaxation
population}. A compromise then is to extract the values of the
cross sections by means of the width of the highest (isolated)
Raman active mode. A rule of thumb is that the lower the
temperature and the higher the energy of the mode the more the
linewidth will be dominated by population relaxation. We can then
use one mode to estimate its $\sigma_{S}$, and all the others
follow automatically from this assumption. The cross sections are
now consistent by construction and the method provides, in fact,
the means to estimate the relative lifetimes of the modes. We call
this procedure (i.e. estimating the $\tau$ of the Raman active
mode with the largest Raman shift, and making the other
cross-sections consistent with this value through the relative
integrated intensities)  the {\it corrected lifetime method}
(CLM).}

\item{If we are comparing different substrates with the same
analyte we could also directly compare the values of
$\tau\sigma_S$ without any further assumption. If there is no
reason to believe that relaxation population is different between
the two substrates, {\it the ratio of $\tau\sigma_S$ for two
different substrates provides a direct comparison of SERS cross
sections}.}
\end{itemize}

\subsection{The asymmetry factor}

\begin{itemize}

\item Problems with measuring the asymmetry factor ``$A$'' can be
as important as those to obtain a reliable lifetime. Heating in
the thermally-dominated regime can be a {\it major} problem. As
demonstrated in Sec. II B, heating can be in part a source for a
larger than normal $A$, if it is not properly identified.

\item Ideally, the best estimation of $A$ can come from
measurements at RT with long integration times and very low
powers. The obtained $A$'s should be more or less the same for a
given analyte in the presence of the same metal (they are mainly
related to the local interaction of the dye with the
metal\cite{MaherResonance}).

\item The product $A \tau \sigma_S$ is measured with high accuracy
in the pumping dominated ``plateau'' at low temperatures.
Accordingly, the best use of the technique is to use one
well-characterized analyte (without photo-bleaching if possible)
as a means to compare the relative SERS cross-sections of the dye
on different substrates of the same metal, by simple comparison of
the plateaus. The product $A \tau \sigma_S$ is perhaps the best
``objective'' comparison between the cross sections of two
substrates without any assumption. However, it is also important
to note that the cross sections we are comparing are not the
average cross sections but rather a biased average towards the
highest cross sections in the sample, as explained in the next
subsection.

\end{itemize}

\subsection{Which cross section do we really measure?}

There are additional complications with the estimation of the
cross section. It is generally accepted that SERS signals are
dominated by the presence of ``hot-spots'' or places with high
local enhancements. The formulae presented above made implicitly
the assumption that the cross section is the same for all $N$
molecules in the sample. The question we want to address now is:
do we obtain an estimate of the average cross section with this
method? We shall show in what follows that what we actually obtain
from the experiment is an estimate which is heavily biased towards
the sites with the highest enhancements; i.e. the method provides
an estimate of the enhancements in hot-spots. This is of course
an advantage and a disadvantage at the same time. We show this
explicitly in this section.

We consider a number of molecules ($i$ from 1 to $N$) in the
sample. Each molecule can have a different anti-Stokes,
($\sigma_{aS}^i$) and Stokes ($\sigma_{S}^i$) SERS cross sections
(according to the asymmetry factor $A^i= \sigma_{aS}^i/\sigma_S^i$
for that molecule). We assume for simplicity that the incident
power $I_L$ is the same for all. The rate equation for the average
phonon population $n^i$ of a vibrational mode of {\it one
molecule}, $i$, and its stationary state solution are given by
expressions similar to Eqs. (\ref{rateEqn}) and (\ref{nSteady}):
Following the other formulae in the previous sections, the Stokes
signal for this molecule is simply $I_S^i=\sigma_S^i I_L$, while
the anti-Stokes signal is given by: $I_{aS}^i= n^i \sigma_{aS}^i
I_L$. The anti-Stokes/Stokes ratio {\it for this molecule},
$\rho^i$, is given by an expression similar to Eq.
(\ref{pumpRatio}). If all the molecules experienced the same
enhancements, then the total Stokes and Anti-stokes intensities
would be $I_S=N I_S^i$ and $I_{aS}=N I_{aS}^i$, and the ratio
would be $\rho=\rho_i$. However, if the molecules have different
cross sections (SERS enhancements), which is a much more realistic
assumption, we then have:
\begin{eqnarray}
I_S &=& \sum_{i=1}^N \sigma_S^i I_L \\
 &=& N \langle\sigma_S\rangle I_L,
\end{eqnarray}
and
\begin{eqnarray}
I_{aS} &=& \sum_{i=1}^N \sigma_{aS}^i I_L \left(\frac{\tau
\sigma_S^i
I_L}{\hbar\omega_L}+e^{-\hbar\omega_v/k_B T}\right) \\
&=& N \langle\sigma_{aS}\sigma_S\rangle I_L \frac{\tau
I_L}{\hbar\omega_L} + N \langle\sigma_{aS}\rangle I_L
e^{-\hbar\omega_v/k_B T}\nonumber.
\end{eqnarray}
All the averages here $\langle\ldots\rangle$ are the usual
statistical ensemble averages.  The aS/S-ratio is then:
\begin{equation}
\rho= I_{aS}/I_S =
\frac{\langle\sigma_{aS}\rangle}{\langle\sigma_S\rangle} \left[
\frac{\langle \sigma_{aS}
\sigma_S\rangle}{\langle\sigma_{aS}\rangle} \tau
\frac{I_L}{\hbar\omega_L}+ e^{-\hbar\omega_v/k_B T} \right].
\end{equation}

Comparing this expression with that obtained previously in Eq.
(\ref{pumpRatio}), what we called $A$ is now replaced by:
\begin{equation}
A^E=\frac{\langle\sigma_{aS}\rangle}{\langle\sigma_{S}\rangle}.
\end{equation}
Note that this is not exactly the average of the $A^i$'s, i.e.
$\langle A \rangle$. Moreover, what we measure (instead of $\tau
\sigma_S$) is now:
\begin{equation} \tau \sigma_S^E=\tau \frac{\langle \sigma_{aS}
\sigma_{S}\rangle}{\langle \sigma_{aS}\rangle}.
\end{equation}

To understand the meaning of $\sigma_S^E$ ($E$ for Ensemble), we
need to understand the sources of non-uniformity. The main
variation is in the cross-sections, because of the large
differences in SERS enhancements at different locations on the
SERS substrate. We typically have values which are, say,
$10^3-10^6$ larger at hot-spots than in other places. However, for
a given molecule, the ratio $A^i=\sigma_{aS}^i/\sigma_{S}^i$
should not vary as much (at least not by more than one order of
magnitude). In fact, this ratio can in a first approximation be
taken as a constant, which is an intrinsic property of the
adsorbed molecule/metal complex\cite{MaherResonance}. Assuming
that $A^i=A$ is the same for all molecules, we simply get that
$A_E=A$, i.e. the $A$ we measure is the correct one. Moreover,
using
 $\sigma_{aS}^i=A\sigma_S^i$, we can express $\sigma_S^E$ as a function of
$\sigma_S^i$ only as:

\begin{equation}
\sigma_S^E=\frac{\langle\sigma_{S}^2\rangle}{\langle\sigma_{S}\rangle}.
\end{equation}
If the distribution of $\sigma_S$'s were fairly uniform (for
example a Gaussian around an average value with a small standard
deviation) then $\sigma_S^E$ would be a good estimate of the
average of the distribution (slightly overestimated). But this is
far from reality in SERS conditions. A more realistic situation is
to have a small number of molecules at a hot-spot (HS), and a large
number of molecules at non-HS positions. For the sake of argument,
let us demonstrate the effect on this average by considering
$10^3$ molecules with one molecule at a HS and 999 molecules at
places with much lower enhancements. The expression given above
would give:
\begin{equation}
\sigma_S^E=\frac{1/1000 \left( 1 (\sigma_S^{\mathrm{HS}})^2 + 999
(\sigma_S^{\mathrm{NHS}})^2 \right)}{1/1000 \left( 1
(\sigma_S^{\mathrm{HS}}) + 999 (\sigma_S^{\mathrm{NHS}}) \right)}.
\end{equation}
If there is a difference of a few orders of magnitude between the
cross sections  $\sigma_S^{\mathrm{HS}}$ and
$\sigma_S^{\mathrm{NHS}}$, the numerator is then easily dominated
by the one molecule at the HS (due to the square). For the
numerator, it is not as clear-cut, it all depends on how much
stronger the HS is and how many molecules are at non-HS positions.
But overall, the denominator is likely to be dominated by
$\sigma_S^{\mathrm{HS}}$, possibly slightly larger if the second
contribution is not negligible. This shows that $\sigma_S^E\approx
\sigma_S^{\mathrm{HS}}$, or perhaps slightly smaller. A more
quantitative argument is only possible if we had a more realistic
distribution of SERS enhancements for a given substrate, but the
qualitative conclusion is that {\it pumping experiments provide a
good lower estimate of the cross-section of the few molecules
experiencing the highest enhancements}.

\subsection{Photo-bleaching}

The last aspect we want to briefly touch upon in this analysis of
cross section estimations via vibrational pumping is
photo-bleaching. The exact mechanisms of photo-bleaching under
SERS conditions are still poorly understood and would deserve a
full study in itself. Photo-bleaching is particularly important in the method we are describing here because molecules sited at HS's (i.e. those exposed to the largest enhancements) are affected to a greater degree. At high power densities, for example, the
population of molecules at HS's will be bleached at a faster rate than the average and this will
change the measured cross section. Studying $\rho$ under
photo-bleaching conditions could, in principle, tell us something
about this distribution. There is robust experimental evidence
that photo-bleaching can complicate the analysis, produce
experimental artifacts, and -to a large extent- even decide the
answer we obtain from a specific experiment. This is added to
several experimental complications in aS/S-ratios which include
the fact that sometimes (depending on the dispersion of the
spectrometer) the anti-Stokes and Stokes sides cannot be measured
simultaneously. This implies a delay between the two measurements
and therefore a difference in exposure times to the laser. By
choosing a certain power level we are, in a way, selecting the
population of HS's that we are going to measure. We shall come
back to this problem in the discussion of the results. A rule of
thumb is that larger laser spots with low power densities (but
sufficient to produce detectable pumping), short integration
times, and non-resonant (with the dye) laser excitation, are in
general preferable for a more reliable estimate of SERS cross
sections. All possible measures should be taken (including the
type of scanning used) to address and minimize the undesirable
effects of photo-bleaching.

\section{Experimental}

\begin{figure}
 \centering{
  \includegraphics[width=10cm, height = 12cm]{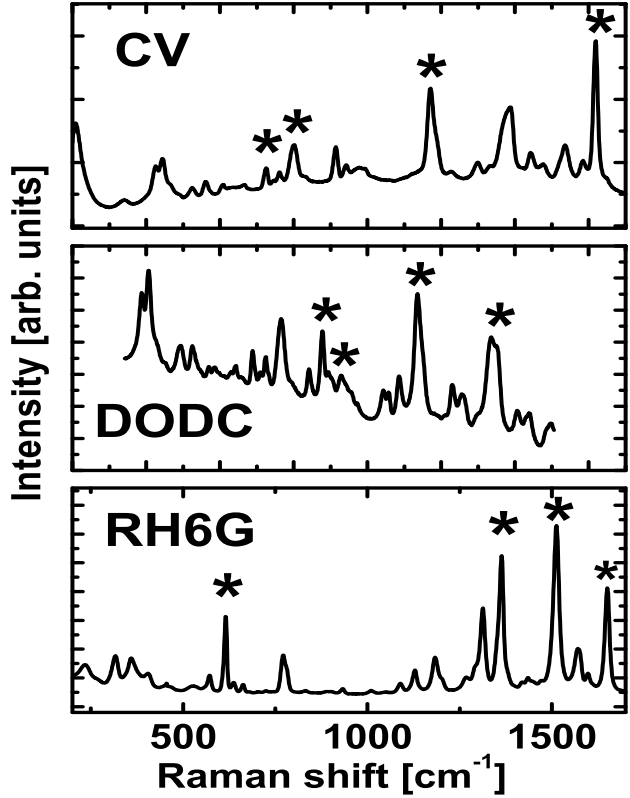}
 }
\caption{SERS spectra of the three analytes used in this paper
taken with a $633\, {\rm nm}$ laser: crystal violet (CV, top),
3,3'-diethyloxadicarbocyanine (DODC, center), and rhodamine 6G
(RH6G, bottom). The temperature dependence of the aS/S-ratios of
the peaks labeled with ``$\star$'' are explicitly shown in Fig.
\ref{TDeps}. Table \ref{tab:FitSigmasAll} includes, in addition,
estimations of cross sections for additional peaks not labeled
here.}
 \label{analytes}
\end{figure}

We turn now to an experimental demonstration of the principles
underlined above. SERS measurements have been performed on
substrates formed by dried Ag colloids on silicon. The colloids
were prepared using the standard Lee and Meisel
technique\cite{Miesel}. SERS active samples were prepared by
mixing the colloids with a $20\, {\rm mM}$ KCl solution in equal
amounts. The analytes were then added to give a concentration of
$10^{-6}\, {\rm M}$ in each case. A small amount of this solution
was then dried on to a silicon substrate. The investigated
analytes are rhodamine 6G (RH6G), crystal violet (CV), and
3,3'-diethyloxadicarbocyanine (DODC). Figure \ref{analytes} shows
the basic SERS spectra of the analytes. We shall show the
explicit temperature dependencies of the aS/S-ratios for only a
few modes (labeled in Fig. \ref{analytes}) while the tables show
additional data for other modes. Samples were mounted in a
closed-cycle He-cryostat (CTI-Cryogenics) with temperature control
in the range $10-300\, {\rm K}$. Raman measurements were performed
using several laser lines of a Kr$^+$ and Ar$^+$-ion lasers which
were focused to a 20 $\mu$m diameter spot. The signal was
collected using a high-numerical aperture photographic zoom lens
(Canon, $\times$10 magnification) onto the entrance slit of a
high-dispersion double-additive U1000 Jobin-Yvon spectrometer
coupled to a liquid N$_2$-cooled CCD detector. Peaks were analyzed
using standard Voigt functions with subtracted backgrounds.  The
variation in the aS/S-ratio with $T$ for each mode was then fitted
to Eq. \ref{pumpRatio}. For this purpose, it is convenient to
modify the expression to:

\begin{equation}
\ln(\rho) =  a + \ln\left[b + e^{-\hbar \omega_{\nu}/k_BT}\right],
\label{lnPumpRatio}
\end{equation}
where $a=\ln(A)$, and $b=\tau\sigma_S I_L/\hbar\omega_L$ which are
the (dimensionless) fitted parameters. The $\tau$'s of the peaks
(necessary to obtain the cross sections) are estimated in two
different ways following the prescriptions in Sec. III B.

\begin{figure}
 \centering{
  \includegraphics[width=12cm, height = 12cm]{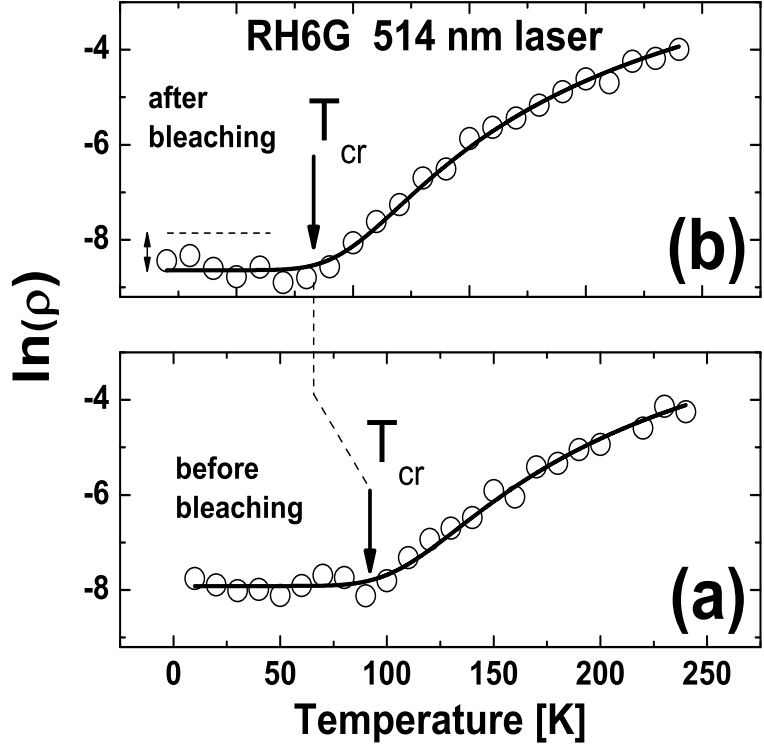}
 }
\caption{Effect of photo-bleaching on the temperature dependence
of the aS/S-ratio at $514\, {\rm nm}$ excitation. The measurement
in (a) is performed at low power density ($25\, {\rm mW}$ on a
$20\, \mu$m spot in diameter). The sample is then exposed to $400
\, {\rm mW}$ for $3\, {\rm min}$ on the same spot and the
measurement is repeated afterwards with the same original low
power density. The decrease in the plateau of $\ln(\rho)$ to
smaller values (and the associated shift of $T_{cr}$ to lower
temperatures) in (b) is consistent with a bleaching of the
molecules on the HS's with the highest enhancement in the first
measurement. The dashed line in (b) is the value at which
$\ln(\rho)$ tails in (a). By choosing a certain power level, we
are selecting the population of HS's that will survive through the
experiment. We are therefore measuring a convoluted property of
the substrate and the photo-stability of the probe.}
 \label{photobleach}
\end{figure}

As discussed earlier, it is of critical importance that the sample
remains stable over the entire length of the experiment. A range
of integration times were used depending on the specific sample
and measurement to achieve a good signal to noise ratio for
different peaks, but power densities were kept to a minimum
compatible with the observation of pumping and exposure to the
beam was minimized as much as possible in between changes of $T$.
We used $\sim 30-50\, {\rm mW}$ spread over the 20 $\mu$m diameter
spot. As the rate of photo-bleaching may change with
$T$\cite{2005PieczonkaCPC} a suitable power density to ensure
reasonable stability must be decided at RT, where the effect is
greatest. The relatively large spot and low power densities have
the triple advantage of ($i$) reducing photo-bleaching to a
negligible level, ($ii$) reducing any indirect laser heating
effects and ($iii$) improving the averaging over cluster
geometries. Still, even under low power density excitation, the
choice of laser line can have an important effect on the
measurement. An explicit example is shown in Fig.
\ref{photobleach} for RH6G under $514 \, {\rm nm}$ excitation. As
explained in the caption, this observation is compatible with a
bleaching of the HS's with largest cross sections. Laser lines
that are strongly resonant with the probe
molecule\cite{MaherResonance} should be avoided so that the
photo-stability of the analyte and the accumulated exposure to the
laser do not become an issue in the interpretation and analysis of
the data. For that reason, we only concentrate hereafter on
results obtained with the $647$ and $676\, {\rm nm}$ lines of a
Kr$^{+}$-laser. These two lines are close enough to the visible
(to profit fully from SERS enhancements) but, by the same token,
they do not produce excessive photo-bleaching for the dyes under
consideration here.

\section{Results and discussion}

\begin{figure}
 \centering{
  \includegraphics[width=18cm, height = 13cm]{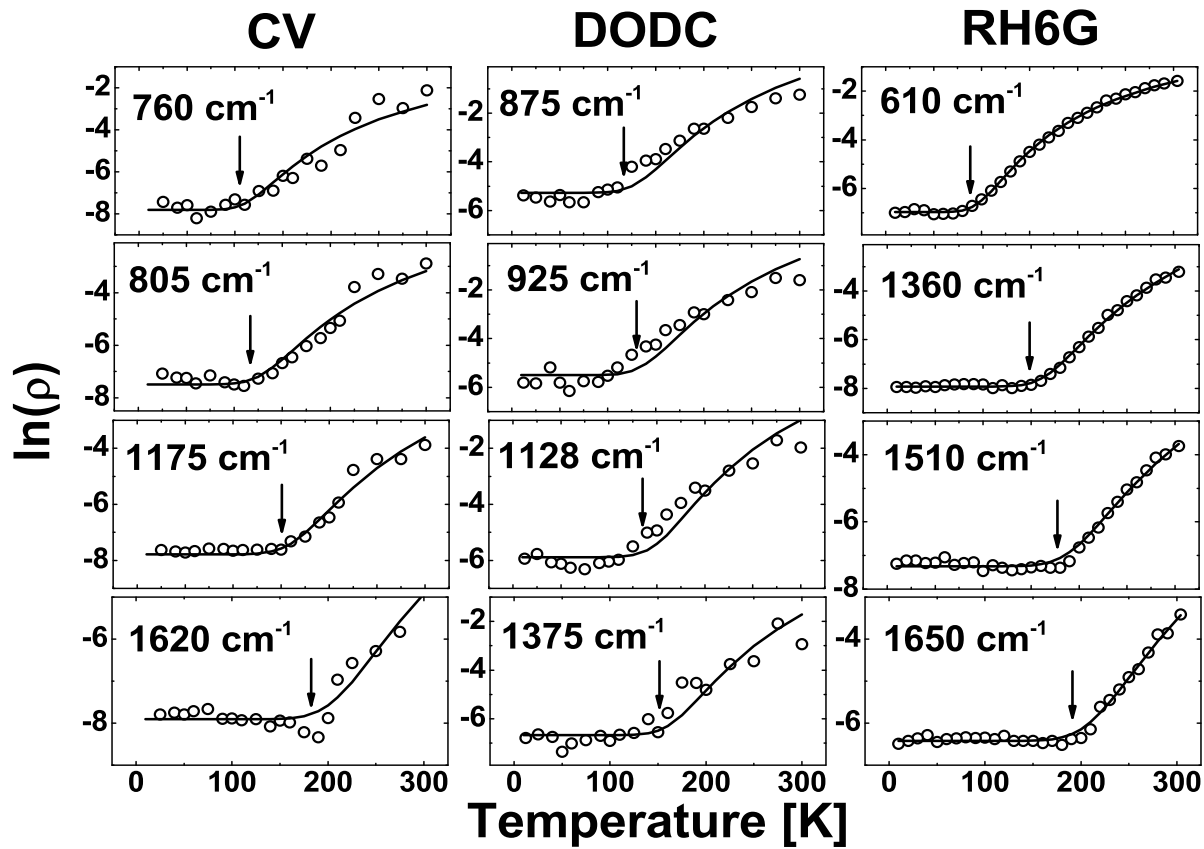}
 }
\caption{Anti-Stokes/Stokes ratios (ln($\rho$)) as a function of
temperature for four modes of each of the analytes investigated
here (all taken using $676\, {\rm nm}$ excitation). The arrows
indicate the crossover points between the thermally-dominated and
pumping-dominated regimes.  Note that the cross over occurs at
different temperatures for different modes, being at higher $T$'s
for larger vibrational energies.  The solid lines represent the
best fit to the experimental data using Eq. \ref{lnPumpRatio}.
Table \ref{tab:FitSigmasAll} presents an analysis of these data
using different assumptions for the lifetimes.}
 \label{TDeps}
\end{figure}

Having all the theoretical tools from the previous sections, we
can now easily scan through several experimental results that
demonstrate the method in practice. Figure \ref{TDeps} shows the
aS/S-ratios (ln($\rho$)) vs. $T$, as measured using the $676\,
{\rm nm}$ line, for four modes of each of the investigated
analytes. The solid lines represent the best fit to the
experimental values using Eq. \ref{lnPumpRatio}. A series of small
imperfections can be seen in the data for some modes, but overall
the behavior is very well represented by Eq. \ref{lnPumpRatio},
with RH6G being the best example. The arrows indicate the
cross-over points $(T_{cr})$ between the thermally-dominated and
pumping-dominated regimes for all the investigated modes. $T_{cr}$
occurs at higher $T$'s for higher energy modes, as expected. The
fact that pumping is observed for each of the modes indicates that
the effect is fairly general and should be observable for a large
number of analytes for as long as $\sigma_S$ is high enough to
make pumping observable at moderate power densities\cite{adenine}.

In order to transform these data into estimates of the cross
sections we need to address the different methods of estimating
$\tau$. Table \ref{tab:FitSigmas} shows the experimental results
for RH6G taken with the $676\, {\rm nm}$ line. For the sake of
comparison, we display the cross section by using two different
estimations of the $\tau$'s. This gives a better idea of how
reliable these numbers are and how differences among them should
be interpreted. Table \ref{tab:FitSigmas} shows both: $(i)$ the
extraction of $\tau$ directly from the widths via $\tau\sim
\hbar/\Gamma$ (this normally renders cross sections which are not
fully consistent with the relative integrated intensities of the
peaks), and $(ii)$ the extraction of $\tau$ from the highest Raman
active mode and the subsequent adjustment of the other
cross-sections to agree with the relative integrated intensities
(as described in Sec. III B). A column with relative intensities
and cross-sections among modes is also provided for completeness.
The anti-Stokes cross sections are obtained through the Stokes
ones via the fitted value of the asymmetry parameter $A$ from Eq.
\ref{lnPumpRatio}. These results prove the point that the
estimation of the lifetime is a crucial step in this method to
transform the experimental values into an estimation of the cross
sections themselves. The values should be interpreted within this
assumption, and their validity should be assessed on a
case-by-case basis. The preferred (more consistent) method is of
course the one that ensures the relative cross sections agree with
the relative intensities of peaks (CLM in Sec. III B).

Finally, we can discuss the effect of estimating the cross
sections for two different excitation wavelengths. The experiment
is somewhat restricted in the choice of excitations (to be
compatible with photo-stability) but still provides a hint of
resonance contributions to the cross sections with some
limitations. Table \ref{tab:FitSigmasAll} shows the cross-sections
obtained for RH6G, DODC, and CV for 647 and $676\, {\rm nm}$
excitations for different modes. All the cross sections were
obtained by the CLM analysis. It is worth noting that in this
latter method we only need (in principle) one temperature
dependent aS/S-ratio for the highest Raman active mode (judged to
be dominated by population relaxation). This together with the
plain SERS spectrum of the dye is enough to estimate all the cross
sections of the modes via the relative integrated intensities. The
results will be consistent (by construction) but not necessarily
accurate (depending on the one mode that has been chosen to
estimate the initial cross section). The main conclusions of these
experiments can be summarized as follows: $(i)$ RH6G shows larger
cross-sections at 647 than at $676\, {\rm nm}$. This is consistent
with a resonant increase known to exist in RH6G towards the yellow
part of the spectrum\cite{MaherResonance}. $(ii)$ On the contrary,
both DODC and CV experience larger cross-sections at $676\, {\rm
nm}$, consistent with the maximum of the absorption of these dyes
being at longer wavelengths with respect to RH6G and, therefore,
having a much more red-shifted resonant interaction than
RH6G\cite{MaherResonance}. $(iii)$ Although there is some
variations from mode to mode, it is generally true that CV
experiences the largest cross-sections in this near-IR region.

The numbers in tables \ref{tab:FitSigmas} and
\ref{tab:FitSigmasAll} deserve a special comment regarding their
interpretation. For a start, the main source of error in these
numbers (as pointed out in the caption of table
\ref{tab:FitSigmas}) is {\it systematic} rather than statistical,
and comes from the different methods used in the estimation of
$\tau$. Error bars for these systematic errors are not easy to
obtain from the data, for they depend on the relative contribution
to the FWHM of population relaxation and dephasing, which is
unknown and changes from mode to mode. The number of significant
figures is decided then by the range in which these numbers span
for a given group of modes and laser excitation. It only makes
sense to compare statistical errors (from the propagation of
errors in the parameters of the fits and the experimental
magnitudes) for two cross sections with the same systematic
errors. This is the situation, for example, of the comparison
between the method presented in this paper and the
power-dependence method (quadratic dependence of the $I_{aS}$
signal as a function of power) as reported in Ref.
\cite{RobJPCB2}. In this latter case the same systematic error
coming from the estimation of $\tau$ affects both results and,
hence, the statistical error bars are meaningful to claim
consistency or inconsistency between the two values. In any other
case, the overriding source of error in these estimates of the
cross sections is systematic.

The order of magnitude of these numbers deserves a special comment
too. According to the results in Sec. III D, we expect these
numbers to represent something close to the maximum available
enhancement in these samples. In addition, the fact that $\tau$
could be in general underestimated by any of the methods presented
here leads automatically to an overestimation of $\sigma$. Last
but not least, the values of $\sigma_{aS}$ are obtained from those
of $\sigma_{S}$ via the asymmetry factor $A$. As a consequence of
all these, we expect the values to be: $(i)$ At the higher-end of
what is known for HS's in SERS for $\sigma_{S}$ and, $(ii)$ with a
larger error for the anti-Stokes values which use the estimation
of the asymmetry parameter $(A)$ two times; once to obtain
$\sigma_{S}$ itself through the fit of $\ln(\rho)$ vs. $T$, and
twice to obtain $\sigma_{aS}$ from $\sigma_{S}$. Indeed, Table
\ref{tab:FitSigmasAll} shows that cross-sections are at the
higher-end of what is commonly accepted for SERS enhancements.
$\sigma_{aS}$ is bound to be less exact than $\sigma_{S}$ because
it depends crucially on the asymmetry parameter and its associated
errors (treated in Sec. III C). The most consistent set of values
are obtained for $\sigma_{S}$ using the CLM-analysis (which should
be the preferred option) and are all of the order of $\sim
10^{-15}\, {\rm cm}^{2}$. In exceptional cases, cross sections of
the order of $\sim 10^{-14}\, {\rm cm}^{2}$ are obtained, which
are still expected if we take into account that the estimation is
for HS's and that the values are possibly overestimated by the
choice of $\tau$. Unrealistically large values of $\sigma_{S}$
beyond that should be interpreted as a limitation of the
technique, in general, and should be tracked down to a problem in
the assignment of $\tau$ for that mode. Even for these cases, it
is still possible to compare $\tau\sigma_{S}$ (or $A \tau
\sigma_S$) for the same modes in different substrates providing a
relative comparison of cross section performances at HS's without
any additional assumptions. This comparison is independent of the
number of molecules involved in the signal, due to the
self-normalizing nature of the method. Table
\ref{tab:FitSigmasAll} summarizes, therefore, a typical example
where the performance of different dyes can be tested through
appropriate estimates of the cross sections derived from the
method presented in this paper.

\section{Conclusions}
We have presented an in-depth discussion of the state-of-the-art
of SERS cross section estimation via vibrational pumping and have
given several experimental examples of it. Despite the
approximations needed and the intrinsic problems of the technique,
it is still an excellent tool (and in many cases possibly the only
tool) to extract an estimation of the SERS cross-sections in
situations where the number of molecules and hot-spots present in
the sample are not known or are difficult to estimate. The {\it
self-normalizing} nature of the method (with respect to the number
of molecules involved) is certainly a major advantage. In
addition, we feel that these latter experimental and theoretical
developments have moved forward to establish vibrational pumping
as a real phenomenon in SERS after almost a decade of controversy
about its very existence. One outstanding issue that surely
deserves further investigation is the effect of photo-bleaching
under SERS conditions (in particular in HS). As we showed in Sec.
III D, the fact that this technique produces an estimate of the
cross sections which is heavily biased towards HS's provides a
unique opportunity to explore their characteristics. This is
particularly true in the comparison among different substrates (in
which the photo-stability of the analyte plays a special role). A
forthcoming study on the topic is in preparation and will be
published elsewhere.

\section{Acknowlegements}
PGE and LFC acknowledge support by EPSRC (UK) under grant
GR/T06124. RCM acknowledges partial support from the National
Physical Laboratory (UK) and the hospitality of the MacDiarmid
Institute at Victoria University (New Zealand) where the
measurements have been performed.  Special thanks are given to Dominik Hangleiter (Wellington College, NZ) for taking the data in Fig. 3.

\newpage

\newpage

\newpage

\newpage

\begin{table*}
    \centering
        \begin{tabular}{|c|c|c|c|c|c|c|c|c|c|c|c|}
          \hline
              & & \multicolumn{5}{|c|}{Widths} & \multicolumn{5}{|c|}{Corrected} \\
          \hline
            Mode & b  & $\tau$ & Rel. $I_S$ & Rel. $\sigma_S$ & $\sigma_S$ (cm$^2$)  & $\sigma_{aS}$ (cm$^2$) & $\tau$ & Rel. $I_S$ & Rel. $\sigma_S$ & $\sigma_S$ (cm$^2$) & $\sigma_{aS}$ (cm$^2$) \\
            (cm$^{-1}$) & ($\times 10^{-5}$) & (ps) & & & ($\times 10^{-15}$) & ($\times 10^{-15}$) & (ps) & & & ($\times 10^{-15}$) & ($\times 10^{-15}$) \\
            \hline
            610 & 25.6 &  0.79  & 1.49 & 4.86 & 6.0 & 22 &  2.6  & 1.49 & 1.49  & 1.8 & 7 \\
            \hline
            780 & 6.85 &  0.22  & 1.82 & 4.75 &  5.8  & 28 &  0.57  & 1.82 & 1.82  &  2.2  &   11   \\
            \hline
            1360 & 1.32 &  0.47  & 1.72 & 0.42 & 0.5  &   14 &  0.12   & 1.72 & 1.72   & 2.1 & 58    \\
            \hline
            1510  & 2.13 &  0.29  & 2.98 & 1.10 & 0.9  &   27 &  0.11   & 2.98 & 2.98  & 3.7  &   114   \\
            \hline
            1650 & 2.11 &  0.32   & 1.00 & 1.00 & 1.2  & 95 &  0.32   & 1.00 & 1.00 & 1.2 & 95 \\
            \hline

        \end{tabular}
    \caption{Values of $b=\tau\sigma_SI_L/\hbar\omega_L$, $\tau$, relative integrated intensities, and cross-sections of the Stokes modes. The values of $\sigma_S$ and $\sigma_{aS}$
    are shown for two analysis schemes - $(i)$ $\tau$'s obtained from the FWHM of the peaks (ignoring dephasing contributions),
    and $(ii)$ corrected $\tau$'s to account for the relative integrated intensities. All data are for RH6G using the $676\, {\rm nm}$
laser line as excitation. The ``$Rel.~I_{S}$'' column is obtained
directly from the relative integrated intensities of the peaks in
the SERS spectrum. Cross sections for the different estimates of
$\tau$ are rounded to one significant figure decided by the range
of variation of $\sigma$ over the modes. We do not quote error
bars because the overriding cause of error in these numbers is
{\it systematic} (rather than statistical) and comes from the
estimation of $\tau$ itself. The systematic errors are not easily
obtainable from the data, for the exact amount of dephasing
contribution to the broadening in each peak is unknown.}
    \label{tab:FitSigmas}
\end{table*}

\begin{table*}
    \centering
        \begin{tabular}{|c|c|c|c|c|c|c|}
          \hline
              & \multicolumn{3}{|c|}{647 nm} & \multicolumn{3}{|c|}{676 nm} \\
          \hline
            Mode  & $\sigma_S$ (cm$^2$)& $\sigma_{aS}$ (cm$^2$)& b  & $\sigma_S$ (cm$^2$)& $\sigma_{aS}$ (cm$^2$)& b \\
            (cm$^{-1}$)&($\times 10^{-15}$)& ($\times 10^{-15}$) & ($\times 10^{-5}$) & ($\times 10^{-15}$) &($\times 10^{-15}$) &($\times 10^{-5}$) \\
            \hline
            \multicolumn{7}{|c|}{RH6G} \\

            \hline
            610 & 20  &  112  &  44.8   & 1.8 & 7  & 25.6  \\
            \hline
            780 & 25   &  136  &  25.9 &  2.2 & 11  &   6.85  \\
            \hline
            1360  &  2  &  463  &  1.26 & 2.1 & 58  &  1.32  \\
            \hline
            1510 & 4 &  499  &  6.38 &  3.7 &   114  &   2.13  \\
            \hline
            1650 & 14  &  232  &  22.5  &  1.2 &   95  &   2.11  \\
            \hline

            \multicolumn{7}{|c|}{DODC} \\

            \hline
            875 & 1.8   &   26  &  11.1  & 0.90  &   33  &   3.04  \\
            \hline
            925 &  1.7  &    34  &  8.40 & 0.94  &   33  &   3.34  \\
            \hline
            970  & 1.7   &  34   &  7.32 & 1.14  &   55  &   2.17  \\
            \hline
            1128 &  2.5  &    51  &  4.93  & 1.74  &   143   &   8.23   \\
            \hline
            1375 &  1.3  &  25  &  1.21  &  1.01  &   132   &   0.30 \\
            \hline

            \multicolumn{7}{|c|}{CV} \\

            \hline
            760  & 1.4   &    1.8    &  93.5  & 0.2  &   0.5  &   53.6  \\
            \hline
            805  & 10.9  &    14.5   &  74.7  & 3.4  &   6.7  &  28.6  \\
            \hline
            910  & 3.9  &   6.5    &  124   & 1.1  &   3.9  &   6.82  \\
            \hline
            1175  & 19.5   &   30.7  &  18.0  &  4.3  &   31.8  &   5.62  \\
            \hline
            1292  & 3.7   &   6.3    &  8.24  & 1.6   &   4.8  &   3.94  \\
            \hline
            1620  & 6.4  &  14.4  &  9.99  & 2.0  &   32.7  &  2.26   \\
            \hline

        \end{tabular}
    \caption{Cross-section for the Stokes and anti-Stokes Raman modes of CV, RH6G and
    DODC extracted from $b=\tau\sigma_SI_L/\hbar\omega_L$ obtained through fitting the experimental data to
    Eq. \ref{lnPumpRatio} for both the 647 and $676\, {\rm nm}$ lasers. Only the lifetime of the mode with the largest
    Raman shift is estimated from its width and the other cross sections are made consistent with this determination via the
    relative integrated intensities (CLM- or ``corrected $\tau$'s'' method in Sec. III B). See the text for further details.}
    \label{tab:FitSigmasAll}
\end{table*}


\begin{thebibliography}{}
\bibitem{adenine} Measurements on adenine with these lasers showed no signs of optical pumping suggesting that the cross-section is small relative to the
analytes used here.
\end{thebibliography}
\end{document}